\documentclass[preprint,aps,prb,groupedaddress,nofootinbib]{revtex4-2}

\usepackage{pgffor}
\usepackage{pdfpages}
\usepackage[utf8]{inputenc}
\usepackage{graphicx}
\usepackage{physics}
\usepackage{xr}
\usepackage{textcomp}
\usepackage{xcolor}
\usepackage[normalem]{ulem}

\makeatletter
\AtBeginDocument{\let\LS@rot\@undefined}
\makeatother

\begin{document}

\title{Relation between interface symmetry and propagation robustness along domain walls based on valley topological photonic crystals}

\author{Gaëtan Lévêque}
\affiliation{Univ. Lille, Institut d'Electronique, de Micro-électronique et de Nanotechnologie \\(IEMN, CNRS-8520), Cité Scientifique, Avenue Poincaré, 59652 Villeneuve d'Ascq, France}
\email{gaetan.leveque@univ-lille.fr}
\author{Pascal Szriftgiser}
\affiliation{Univ. Lille, CNRS, UMR 8523-PhLAM-Physique des Lasers Atomes et Molécules, F-59000 Lille, France}
\author{Alberto Amo}
\affiliation{Univ. Lille, CNRS, UMR 8523-PhLAM-Physique des Lasers Atomes et Molécules, F-59000 Lille, France}
\author{Yan Pennec}
\affiliation{Univ. Lille, Institut d'Electronique, de Micro-électronique et de Nanotechnologie \\(IEMN, CNRS-8520), Cité Scientifique, Avenue Poincaré, 59652 Villeneuve d'Ascq, France}

\date{\today}

\begin{abstract}
Valley photonic crystals provide efficient designs for the routing of light through channels in extremely compact geometries. The topological origin of the robust transport and the specific geometries under which it can take place have been questioned in recent works. In this article, we introduce a design for valley photonic crystals with richer arrangement possibilities than the standard valley photonic crystals based on two holes of different sizes in the unit cell. Our approach is based on the permutation of three sets of rhombi in an hexagonal lattice to investigate the interplay between Berry curvature, valley Chern number and chirality of interfaces to achieve robust edge-modes propagation along domain walls. We study three types of interfaces with different symmetries: the non-chiral interface with glide-mirror symmetry commonly used in honeycomb-type valley crystals, and two chiral interfaces with or without inversion symmetry of the adjacent bulk lattices. In the latter case, no valley topology is expected. We show that for the three families, edges preserving the shape of the interface through 120° sharp corners can sustain edge-modes with comparable robustness. Moreover, interfaces with glide-mirror symmetry offer promising performances in circuits with more exotic configurations, like 60° and 90° corners or arbitrary curves in which valley preservation is not guaranteed. Our work raises questions about the topological origin of the robustness of transport in valley photonic crystals, discusses the role of the chirality of the interfaces in the propagation around sharp corners, and provides a lattice scheme with broad design possibilities.
\end{abstract}

\maketitle
\section{Introduction}
Topological photonics relies on mimicking quantum Hall effects in solid-state physics to transport an electromagnetic signal through topologically protected edge-modes \cite{wang_Hybrid_2022, ozawa_Topological_2019, wu_Scheme_2015, parappurath_Direct_2020, barik_Twodimensionally_2016, barczyk_Interplay_2022}, in order to realize compact and low-loss devices in integrated photonics \cite{kumar_Phototunable_2022,arora_Direct_2021,shalaev_Robust_2019,jalalimehrabad_Chiral_2020,barik_Chiral_2020}. A most favored strategy relies on valley topological photonics, where an interface is created between two mirror-image photonic crystals with broken inversion symmetry~\cite{ma_AllSi_2016,he_Silicononinsulator_2019, xue_Topological_2021}. The topological phase transition at the interface ensures the change in sign of valley Chern numbers, at the origin of the existence of topological modes that travel along $\Gamma$K and $\Gamma$K' interfaces. Those edge-modes have been shown to sustain unidirectional propagation with low back-scattering or valley number conversion in a variety of circuits \cite{ma_AllSi_2016,xue_Topological_2021,leveque_scatteringmatrix_2023}, and low sensitivity to certain types of lattice imperfections and defects \cite{orazbayev_Quantitative_2019,arora_Direct_2021}.

However, the topological interpretation of the transport properties of photonic devices faces a major obstacle: no robust topological protection is expected for photons in bi-dimensional systems unless time-reversal symmetry is broken, which is realized in Chern photonic insulators \cite{haldane_Possible_2008,wang_ReflectionFree_2008,wang_Observation_2009}. Without magneto-optic effects, topological properties of photonic bands rely on crystalline symmetries that are broken by the interface or structural defects. Additionally, valley photonics suffers from a lack of bulk-edge correspondence, and valley Chern number $C_v$ is not expected to be a true topological number in the sense that nothing guarantees that it should be an integer or half an integer \cite{xue_Topological_2021,qian_Topology_2018}. In particular, the often discussed value of $C_v=\pm 1/2$ is only reached in the limit of small gaps~\cite{ma_AllSi_2016, qian_Topology_2018}, when the Berry curvature is the most confined around $K/K'$ points. For larger band-gap, the Berry curvature extends far from the $K/K'$ points and the valley Chern number takes arbitrary values, decaying to zero. Employing a honeycomb lattice consisting of two circular holes with different sizes per unit cell, Yang et al \cite{yang_Evolution_2021} showed that if the radius of one hole reduces to zero, the chiral phase vorticity characteristic of the eigenmodes in a lattice with broken inversion symmetry remains, and an interface still sustains robust edge modes even though the Berry curvature and valley Chern numbers reach zero.

The question of the robustness of structural defects has also been addressed recently in several works \cite{orazbayev_Quantitative_2019, arregui_Quantifying_2021}, showing in particular that the long-range protection to small imperfections inherent to micro-/nano-fabrication is not better for linear topological waveguides as compared to trivial waveguides in the slow-light regime \cite{rosiek_Observation_2023}. In addition, the problem of the relation between the exact geometry of the interface and the topological bulk properties is still under question, and it is of important matter for the optimal design and choice of lattices for applications. For instance, it has been noted that for valley photonic crystals of equal bulk geometry, zigzag (face-to-face) interface channels are more robust to backscattering at 120° corners than bearded interfaces and that triangular holes show more protection than circular holes~\cite{mohammed_engineering_2024}. In recent work, Yu et al \cite{yu_Impact_2024} have investigated the modification of edge states at the interface of two valley photonic crystals made of triangular holes with mirror symmetry when the interface is shifted away from the usual zigzag interface. They showed that the interface states changed from an ungapped to a gapped dispersion, which results in much stronger backscattering at obstacles in the latter case. This body of work demonstrates that the existence and the robustness to defects of valley-dependent edge states depend not only on the design of the bulk (in particular through its Berry curvature and valley Chern number) but also on the interface geometry.

In this article, we introduce a design for valley photonic crystals with richer arrangement possibilities than the standard valley photonic crystals based on two holes of different sizes in the unit cell. Our purpose is to enlarge the number of possible interface geometries between different photonic crystals with nontrivial valley topology. In this way, we can provide a systematic study of the presence of interface states and their robustness to sharp corners of different geometries. We propose a versatile design based on a kagome lattice investigated by Vakulenko et al \cite{vakulenko_Field_2021}, where a dielectric membrane is drilled with a periodic arrangement of three rhombi air holes per primitive cell. We adapt this geometry to the valley Hall effect considering a super-cell made of nine rhombi gathered in three sets differing by their size. Permutations of those sets in the lattice do not modify its band structure nor the valley Berry curvature, but allow the creation of three distinct families of bearded interfaces between a lattice and its image through a given permutation. The first family corresponds to interfaces with glide-mirror symmetry, the second to interfaces between lattices differing by a combination of glide-mirror symmetry and a sub-lattice translation, and the last to interfaces between lattices differing only by a sub-lattice translation. We then compare the propagation of interface modes along domain walls with double 120° corners and through triangular resonators, and finally along more exotic edges like 60° corners, 90° corners, and arbitrary curves. This system greatly expands the number of interfaces offered by popular geometries consisting of honeycomb lattices with two units per primitive cell, which can only support two types of crystalline interfaces. It allows probing the interplay of Berry curvature, valley Chern number, and the symmetry of the interface in robust edge-mode propagation. We show that even interfaces without inversion of the valley Chern number can support robust transport through sharp corners. We discuss how the local symmetry of the interface plays a major role in photonic transport along domain walls. Our study is focused on photonic crystals for THz applications, but the results can be extrapolated to other wavelengths by direct geometrical scaling.

\begin{figure*}[t]
	\includegraphics[width=14cm]{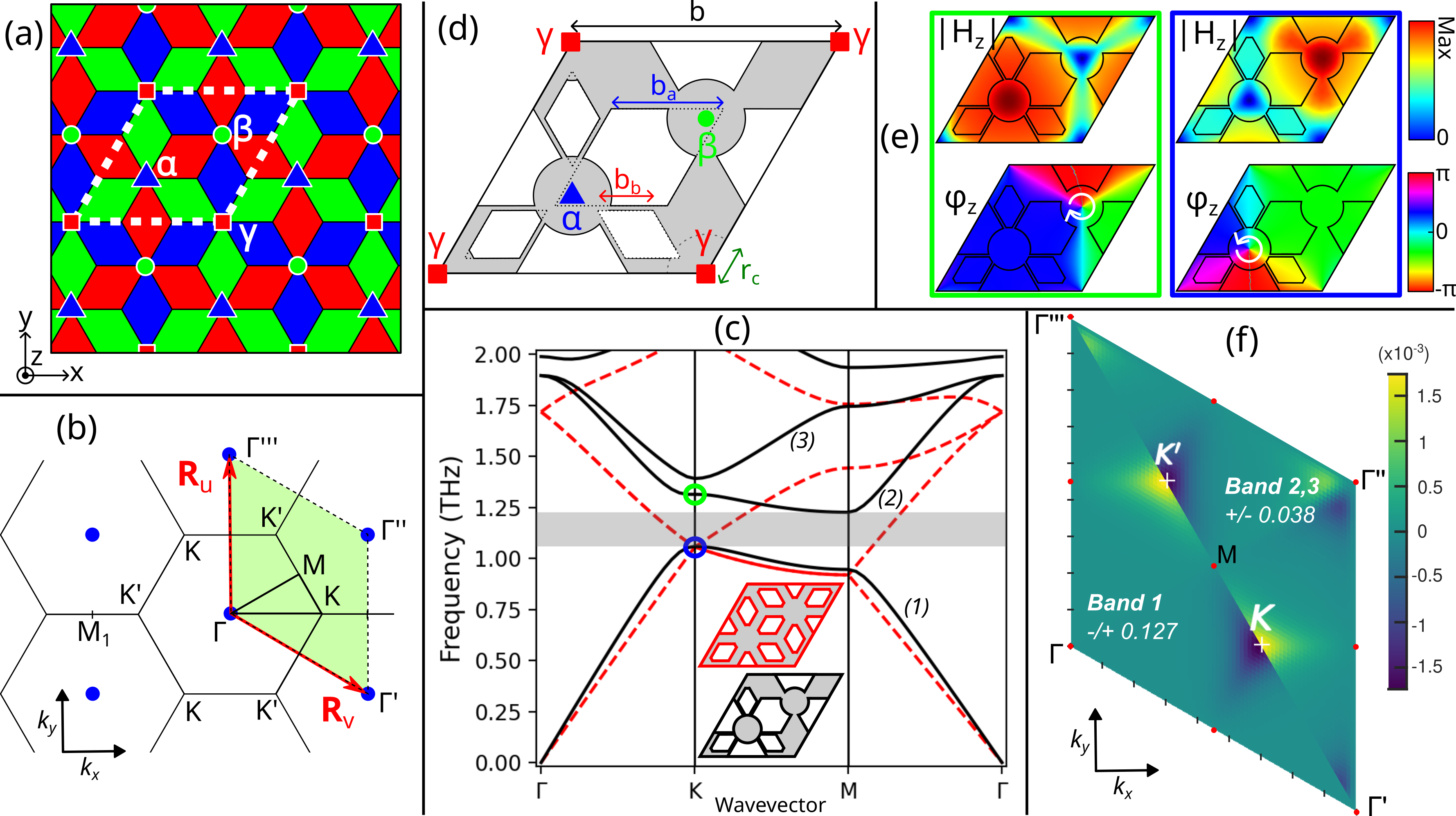}
	\caption{(a) Representation of the lattice, together with the three families of rotation axes associated to each subset of rhombi. (b) Reciprocal lattice and first Brillouin zone in green. (c) Dispersion diagram of the crystal with identical (resp. different) rhombus sizes in dashed red (resp. solid black) lines. (d) Representation of the primitive cell, with air holes in white and solid material in gray. (e) Distributions of $H_z$ amplitude and phases at the $K$ points on both limits of the band-gap. (f) Distribution of the Berry curvature for the first band and degenerate second and third bands in the first Brillouin zone.}
	\label{fig:fig1}
\end{figure*}

\section{Description of the photonic crystal lattice}
The lattice is shown in Fig.~\ref{fig:fig1}(a), with the primitive cell emphasized by a white-dashed line. The full system is composed of three subsets of rhombus air holes, identified by their colors which represent different sizes. The band structure and Berry curvature of the photonic crystal can be modified changing independently the three sizes. A complementary representation of the crystal consists of three nonequivalent sub-lattices of point rotation axes, identified by blue triangles ($\alpha$ sites), green disks ($\beta$ sites), or red squares ($\gamma$ sites). Each point axis with a given color is a $C_6$ axis for the subset of rhombi with the same color, but reduces to a 3-fold symmetry axis for the whole lattice if the three colors are different (i.e., if the three sets of rhombi have a different hole size). With this definition, the spatial arrangement of the colored rhombi is equivalent to the configuration of the three sub-lattices of point rotation axes. Figure \ref{fig:fig1}(b) shows the reciprocal lattice (blue dots) with basis vectors $\vb{R}_u$ and $\vb{R}_v$, together with the chosen first Brillouin zone in green.

The period of the lattice is fixed to $b=73$ µm and the refractive index of the dielectric material is $n=3.5$. We consider only TE polarization, the magnetic field being perpendicular to the $xy$ plane of the lattice. All simulations have been performed with Comsol Multiphysics. We examine first a reference lattice where all rhombi have the same dimensions (mono-color crystal) with an edge length of 14.6 µm, as represented in red on Fig.~\ref{fig:fig1}(c). Its dispersion diagram in red dashed-lines shows a Dirac cone at $K$ points and frequency $F=1.055$ THz. The gap can be opened by taking different dimensions for all three subsets of rhombi. If only one subset has a size different from the two others, the corresponding rotation axis remains 6-fold for the full crystal and the gap does not open. The chosen primitive cell is shown in Fig.~\ref{fig:fig1}(d), and leads to a band-gap of about 160 GHz centered at $F=1.14$ THz. One subset is composed of rhombi with zero size (the hole is filled with dielectric material), the small rhombi have the same dimension as in the reference lattice, $b_a=14.6$ µm, and the large rhombi have an edge $b_b=29.7$ µm. In order to compensate for the fact that the gap tends to close when the total surface of the air increases within the primitive cell, the tips of the rhombi have been truncated using a disk with radius $r_c=10.4$ µm located at each $C_3$ axis. Distributions of the magnetic field amplitude, $\abs{H_z}$, and phase, $\phi_z$, inside the primitive cell are plotted in Fig.~\ref{fig:fig1}(e) at the $K$ point for frequencies corresponding to the upper and lower limits of the band-gap and indicated by the colored circles in Fig.~\ref{fig:fig1}(c). As expected for a valley topological crystal, magnetic-field singularities with opposite phase vorticities are observed for the low- (respectively high-) frequency mode at the $\alpha$ (resp. $\beta$) sites. 

The topological nature of the gap is confirmed by the distribution of the Berry curvature and associated valley Chern numbers (VCN) \cite{blancodepaz_Tutorial_2020,vanderbilt_Berry_2018} shown in Fig.~\ref{fig:fig1}(f). The Berry curvature is plotted for the isolated first band and the degenerate second and third bands considered together. For the first band, the Berry curvature reaches its largest magnitude but with opposite signs at $K$ and $K'$ points, as the Berry curvature is odd under parity symmetry. The same behavior is obtained for the degenerate second and third bands, however with opposite signs as compared to the first band at each $K$ and $K'$ points. The VCNs computed by integrating the Berry curvature over half of the first Brillouin zone centered at the $K'$ point are $C_v=$0.127 for the first band and $C_v=$-0.038 for the degenerated second and third bands, confirming that the lattice exhibits valley topology features.

\section{Families of possible interfaces}
Once the bulk of the lattice has been characterized, we identify the types of interfaces that can be constructed between two lattices with distinct arrangements of rhombi. By arrangement, we mean the different image lattices obtained after any permutation of the rhombi colors or sub-lattices of point axes of an arbitrary original lattice. In section I of the Supplemental Materials we show that among the five possible permutations not equal to identity, the cyclic permutation $(\beta,\gamma,\alpha)$ (resp. $(\gamma,\alpha,\beta)$) corresponds to a sub-lattice translation by any vector of form $\vec{\alpha\gamma}$, $\vec{\beta\alpha}$ or $\vec{\gamma\beta}$ (resp. $\vec{\alpha\beta}$, $\vec{\beta\gamma}$ or $\vec{\gamma\alpha}$). In this case, the original lattice and the image lattice after permutation have identical Berry curvatures and valley Chern numbers. The three last permutations are the transpositions $(\alpha,\gamma,\beta)$, $(\gamma,\beta,\alpha)$ or  $(\beta,\alpha,\gamma)$, with the sub-lattice of axes $\alpha$, $\beta$ or $\gamma$ respectively unchanged. The original and image lattices differ by a parity transformation and a sub-lattice translation. They have Berry curvatures and valley Chern numbers of opposite signs. As a comparison, in standard valley photonic crystals made of a large and a small hole in a honeycomb lattice, only one type of image lattice is possible by permuting the small and large holes of the primitive cell. In this case, the transformation is fully equivalent to a parity operation.

\begin{figure*}[!t]
	\includegraphics[width=17.5cm]{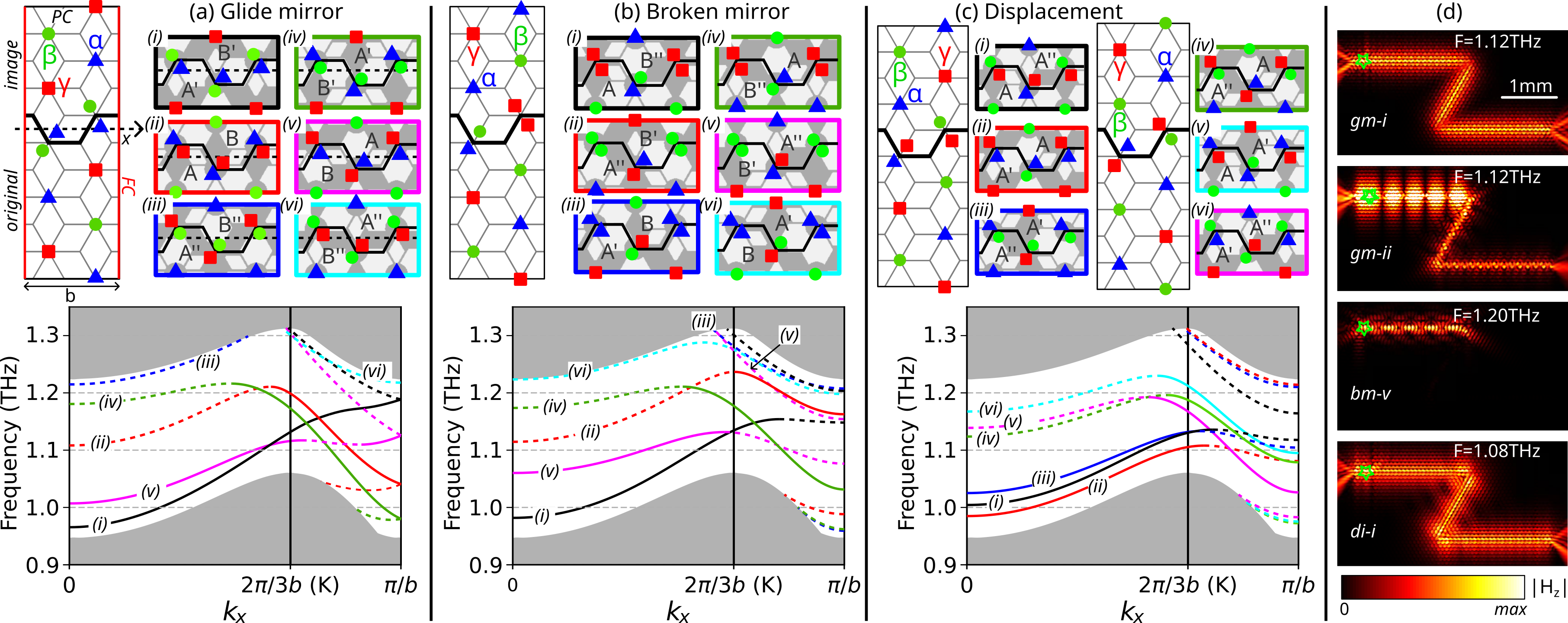}
	\caption{(a) Top: super-cell of a typical interface obtained by transposition with glide-mirror symmetry between original and image lattices, and corresponding six possible interfaces; bottom: dispersion diagrams of the six interfaces. A solid line indicates robust protection against back-reflection at sharp corners, and a dashed line corresponds to strong reflection. (b) Same as (a) but for a transposition breaking the glide-mirror symmetry. (c): Same as (a) but for cyclic permutations. (d) Amplitude profiles of the edge-modes propagating along a Z-interface, for selected frequencies and interfaces.}
	\label{fig:fig2}
\end{figure*}

In this work, we restrict our study to bearded interfaces, which always present a glide-mirror symmetry in honeycomb valley crystals with two-holes per unit-cell. Using this geometrical interface, we will study the properties of interface channels when different permutations of the initial lattice are considered on each side of the interface. The corresponding delimitation between the original and image lattices is represented as a thick black line on the four super-cells in the top panels of Fig.~\ref{fig:fig2}(a-c). The dispersion diagrams of those uni-dimensional interfaces with period $b$ have been simulated by applying Floquet conditions (FC) along the boundaries in red, perpendicular to the interface (see Fig.~\ref{fig:fig2}(a)), and perfect-conductor (PC) conditions along the two remaining horizontal boundaries of the rectangular super-cells.


Depending on the specific permutations between the original and image lattice, we identify three families of interfaces, labeled glide mirror (Fig.~\ref{fig:fig2}(a)), broken mirror (Fig.~\ref{fig:fig2}(b)), and displacement (Fig.~\ref{fig:fig2}(c)). Panels (a) and (b) correspond to cases where the original (top) and image (bottom) lattices differ by a transposition: they are mirror-symmetric (up to a sub-lattice translation) and one of the three sub-lattices of rotation axes is left unchanged. The supercell shown in the top-left part of panel (a) presents a glide-mirror symmetry materialized by the dashed line: it is the analog of the bearded interface between two symmetric honeycomb lattices with two holes per unit-cells. Six distinct interfaces are obtained by applying the six possible permutations of the rotation axes in the original (and automatically in the image) lattice. They are shown in the six top central panels of Fig.~\ref{fig:fig2}(a). The bottom panel of Fig.~\ref{fig:fig2}(a) displays the calculated dispersion of the interface modes for the six possible permutations shown in the panels above, (\textit{i}) to (\textit{vi}). As expected, due to the glide-mirror symmetry, the dispersion curves are all folded at the edges of the first Brillouin zone of the interface, $k_x=\pm \pi/b$, leading to a degeneracy point. Some of the interface modes traverse completely the bandgap and others do not. We will discuss their differences below.

In the second family of interfaces, corresponding to panel Fig.~\ref{fig:fig2}(b), the original and image lattices are related by a transposition for which the unchanged sub-lattice of rotation axes is on the delimitation of the bearded interface (for example the $\gamma$ axes in the supercell of Fig.~\ref{fig:fig2}(b)). In this situation the glide-mirror symmetry is broken: the image lattice is obtained by the combination of a glide-mirror and a sub-lattice translation. Six distinct "broken-mirror" interfaces are obtained, and all the edge modes are gapped at $k_x=\pm \pi/b$. 

In the last case, panel (c), the original and image lattices are related by a cyclic permutation of the three sub-lattices of rotation axes, equivalent to a translation. There are two families of three interfaces depending on whether the cyclic permutation is in one direction or the other. As the original and image lattices differ only by a translation, those interfaces are called "displacement" edges in the following. Interestingly, and instead of interfaces obtained by transpositions, the band diagrams show that all six interfaces sustain edge-modes, \textcolor{black}{which are also all gapped.}

In the following, glide-mirror, broken-mirror and displacement interfaces will be labeled respectively $gm-x$, $bm-x$ and $di-x$ where $x = i, ..., vi$ is the number of the interface in the corresponding panel of Fig.~\ref{fig:fig2}. Note that broken mirror and displacement interfaces are both chiral: the interface cannot overlap by translation and rotation with its image obtained by any reflection. On the other hand, the glide-mirror interfaces are not chiral.

\section{Study of the robustness of photonic transport around corners}

In the lower panels of Fig.~\ref{fig:fig2}(a)-(c) we have identified in solid lines the portions of each dispersion curve corresponding to robust unidirectional propagation. To do so, we have simulated the propagation through the edge modes along an interface consisting of two consecutive  120° sharp corners, see Fig.~\ref{fig:fig2}(d), and Fig.~S2 of the Supplemental Materials for all interfaces. The modes are excited from the left side of the edge by a localized in-plane electric dipole. The fourteen edge-modes obtained for the glide-mirror, broken-mirror and displacement interfaces, Figs.~\ref{fig:fig2}(a-c), present all a portion of their dispersion curve with robust propagation characterized by low reflection at corners, and identified by solid lines on the dispersion diagrams. This is for example the case for the \textit{gm-i} interface at $F=$1.12 THz, which is represented in the top panel of Fig.~\ref{fig:fig2}(d). The parts of the dispersion curves plotted as a dashed line correspond to edge-modes with noticeable backscattering at the first corner, which occurs in two situations. In the case where a dispersion curve presents an extremum, two modes with equal frequencies but different wave-vectors can be simultaneously excited by the source, with only one of them showing robust propagation. For example, the field distribution of the edge-mode propagating along the \textit{gm-ii} interface at $F=1.12$ THz shows a clear interference pattern between the source and the first sharp corner due to large back-scattering of the low-wavevector edge-mode, while the smoother profile beyond the first corner is attributed to the robust edge-mode with larger wave-vector. In the second situation where the dispersion curve has no extremum, like for the broken-mirror interface \textit{bm-v} at frequencies close to F=1.2 THz, the mode is strongly back-reflected at the first corner.

The robustness of the propagation is expected for edge-modes corresponding to glide-mirror and broken-mirror interfaces, which separate two topological lattices with opposite Berry curvatures and VCNs, and then must obey valley number conservation along the propagation direction. However, the equal robustness of edge-modes along displacement interfaces is more surprising, as both sides of the interface have local VCNs of the same signs. For example, Fig.~\ref{fig:fig2}(d) clearly shows that the mode at $F=1.08$ THz of interface \textit{di-i} propagates without noticeable reflection through both sharp corners. \textcolor{black}{This confirms previous reports in which robust propagation along specific displacement interfaces in triangular photonic crystals has been reported~\cite{yang_Evolution_2021}, even though neither of the two interfaced lattices had any local Berry curvature.}

\begin{figure*}[!t]
	\includegraphics[width=17.5cm]{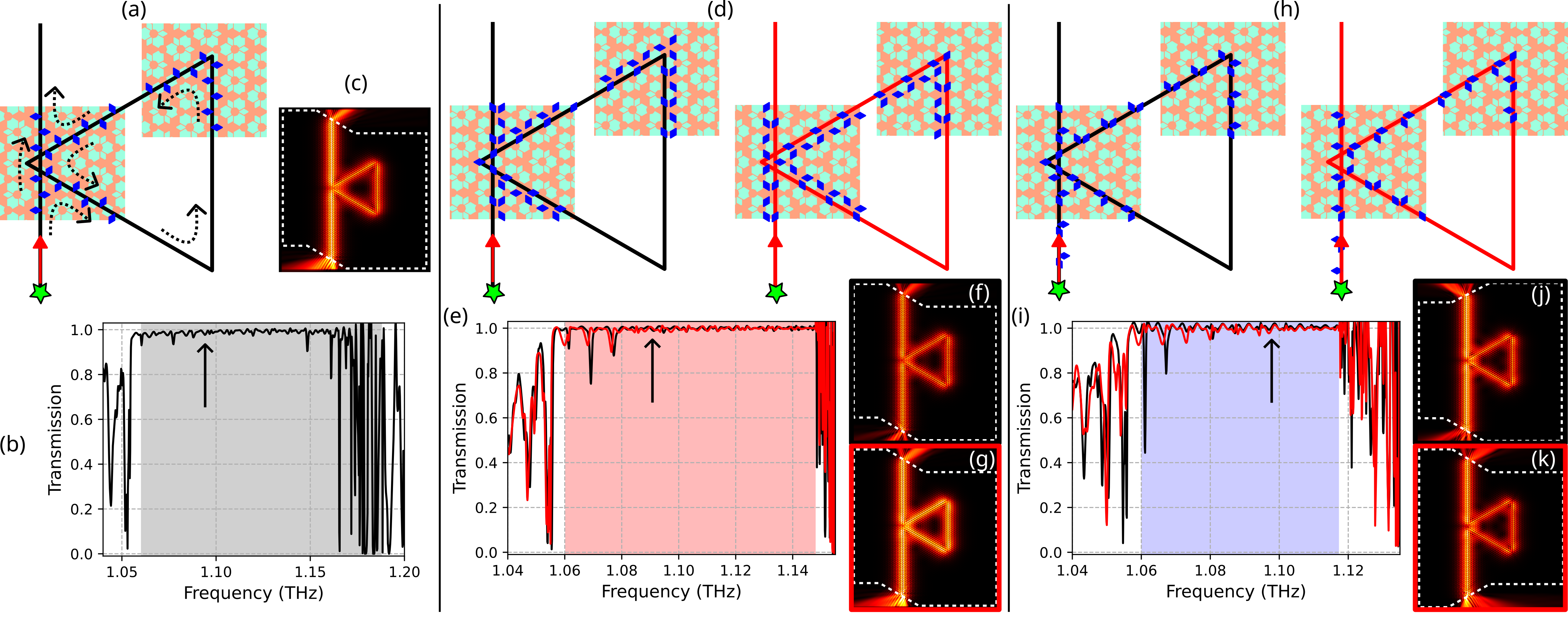}
	\caption{Simulation of the transmission spectra of an edge-mode through a triangular resonator addressed by a straight waveguide. (a) Case of the glide-mirror symmetric interface. The structure of the lattice is plotted around the splitter and the corner, and the small rhombi are shown in blue. The wave is launched along the red arrow from a dipole located at the green star. (b) Transmission spectra, where the colored area indicates the bandwidth of the edge-mode. (c) Amplitude of the magnetic mode for $F=1.092$ THz (black arrow on the dispersion diagram). The white line shows the limit of the lattice. (d-g) Same as (a-c) for the broken-mirror interface. The black and red circuits correspond to interfaces where the original and image lattices of Fig.~\ref{fig:fig2} have been exchanged. (h-k) Same as (d-g) for the displacement interface.}
	\label{fig:fig4}
\end{figure*}

To gain further insights into the scattering properties of the corners, we have simulated the propagation of three types of edge-modes (\textit{gm-i}, \textit{bm-i} and \textit{di-i}) along circuits consisting of a triangular resonator connected to a line interface waveguide, see Fig.~\ref{fig:fig4}. In all three cases, the triangles have a side length of 28 periods. Figures~\ref{fig:fig4}(a,d,h) show the geometry of the three configurations with detailed views of the lattice in the regions of connection between the resonator and the line waveguide and of one corner of the resonator. For better visualization of the different interfaces, the small rhombi in the unit cells along the edge are emphasized in blue. For \textit{bm-i}~(Fig.~\ref{fig:fig4}(d)) and \textit{di-i}~(Fig.~\ref{fig:fig4}(h)), two configurations with exchanged original and image lattices are considered, displayed in black and red in the upper panels. Each of them has a different local configuration at the corners. Below we will analyse their influence on the robustness of transport. 

In the simulations, a mode is excited by a dipolar source located at the green stars. The transmission spectra are shown in Fig.~\ref{fig:fig4}(b,e,i) after normalization by the power flowing along the straight waveguide alone. Typical distributions of the magnetic-field amplitude are plotted for a frequency $F=1.092$ THz in Fig.\ref{fig:fig4}(c,f,g,j,k). The white dashed line indicates the limit of the photonic crystal, fully embedded in the homogeneous dielectric environment. Note that in all simulations, the shape of the boundary between the lattice and the external domain is oblique at both ends of the line waveguide to minimize back-reflections \cite{ma_AllSi_2016}. In such a circuit, an edge-mode with perfect unidirectional propagation would follow the directions indicated by dashed black arrows in the left panel Fig.~\ref{fig:fig4}(a), corresponding to transport along a single valley with conserved forward propagation. As a consequence, the edge-mode cannot be backscattered in the direction of the source and the transmission should be unity. Any observation of resonances in the transmission spectrum should then be related to the breakup of uni-directionality either at the connection between the resonator and the straight waveguide or at the corners of the resonator.

The transmission spectrum for the interface with glide-mirror symmetry, Fig.~\ref{fig:fig4}(b), is flat and very close to unity inside the transmission band of the mode (gray area), except for frequencies beyond 1.17 THz where the dispersion curve tends to flatten in Fig.~\ref{fig:fig2}(a). Shallow resonances can be seen close to 1.06, 1.08 and 1.15 THz, which is the signature of weak back-reflection at the cavity corners or at the connection with the straight waveguide. But, overall, transmission is quasi-unity, a consequence of highly unidirectional propagation in the circuit and nearly perfect conservation of the valley number in the topological interpretation. In accordance, the typical amplitude distribution inside the circuit, Fig.~\ref{fig:fig4}(c), is smooth and exempt from any interference pattern at the input waveguide. 

The second case corresponds to the broken-mirror interface displayed in the middle panel of Fig.~\ref{fig:fig4}. The edge is now chiral \textcolor{black}{in the sense that it does not overlap by rotation with its image by any reflection}. Therefore, two distinct circuits can be constructed: one from a given broken-mirror interface and a second one from its mirror-symmetric with respect to the propagation direction. These two possible circuits are plotted in black and red in Fig.~\ref{fig:fig4}(d). Note that the blue triangles flip from one side of the edge to the other when going from the black to the red configurations, resulting in different local arrangements of the holes at the splitter and corners. The transmission spectra for both circuits are, nevertheless, very similar and close to unity within the transmission band (red area in Fig.~\ref{fig:fig4}(e)), except near 1.07 and 1.08 THz where two shallow resonances are observed. Moreover, transmission minima are deeper for the circuit underlined in black, which shows that the different configurations of corners and splitter resulting from the edge chirality affect the preservation of the edge-modes uni-directionality. Far from those resonances, the transmission is nearly unity and the field distributions are again very smooth, Figs.~\ref{fig:fig4}(f,g).

The last configuration we have considered is displayed in the right panels of Fig.~\ref{fig:fig4}. The circuit is based on a displacement interface. This interface is also chiral and leads to two non-overlapping circuits, Fig.~\ref{fig:fig4}(h). The transmission is also high in the transmission band shown in blue in Fig.~\ref{fig:fig4}(i). Four resonances are visible between 1.06 and 1.08 THz with the transmission minima being lower for the black circuit than for the red one. Beyond 1.08 THz, the transmission spectra is almost unity, no back-scattering occurs as shown by the flat edge-mode field distributions of Fig.~\ref{fig:fig4}(j) and (k).
Those simulations confirm the above observation that a displacement edge, which separates two photonic crystals with equal VCN and, therefore, does not present any valley topological properties, can support propagation around sharp corners as robust as an interface made of two mirror-image crystals with opposite VCNs.

As no topological phase transition occurs at the displacement interface, valley topology cannot explain the robustness of the propagation. We can however observe that, for a chiral interface, an edge-mode travelling along a given direction will not "see" the same environment as an edge-mode travelling along the opposite direction. For instance, on the left circuit of Fig.~\ref{fig:fig4}, the edge-mode emitted by the source sees blue triangles on the right, which is always the case if it propagates along directions indicated by dashed arrows on Fig.~\ref{fig:fig4}(a), allowing unity transmission. If the edge-mode is back-scattered at the splitter or a corner, it propagates along the opposite direction and sees the triangles on its left. \textcolor{black}{Finally, for an edge-mode traveling along an interface, we can arbitrarily attribute an helicity of +1 (resp. -1) to the corresponding direction if the blue diamonds point to the right (resp. left) when oriented by the wave-vector of the edge-mode}. Hence, for this specific situation, robust propagation along a chiral interface can be phenomenologically related to the interface-related helicity conservation instead of bulk-related valley number conservation. 

\section{Transmission around arbitrary corners}
At this point, no qualitative differences seem to appear in the propagation around corners between valley- or helicity-protected edge modes. However, several studies have evidenced that valley topological edge modes have good propagation properties along more exotic shapes of the domain wall such as corners with angles departing from 120°: robust transmission has been demonstrated along 60° and 90° corners, despite the fact that in the first case, different interfaces are connected, and in the second case propagation occurs along $\Gamma M$ directions~\cite{lu_Observation_2017,wong_Gapless_2020,yang_Evolution_2021}. In the following, we investigate how the glide-mirror, broken-mirror and displacement edge modes behave in similar configurations.

As a preamble, we note that there are several ways of connecting two interfaces with 120° or 60° corners in the investigated lattices.
\begin{figure}[h!]
	\includegraphics[width=8.5cm]{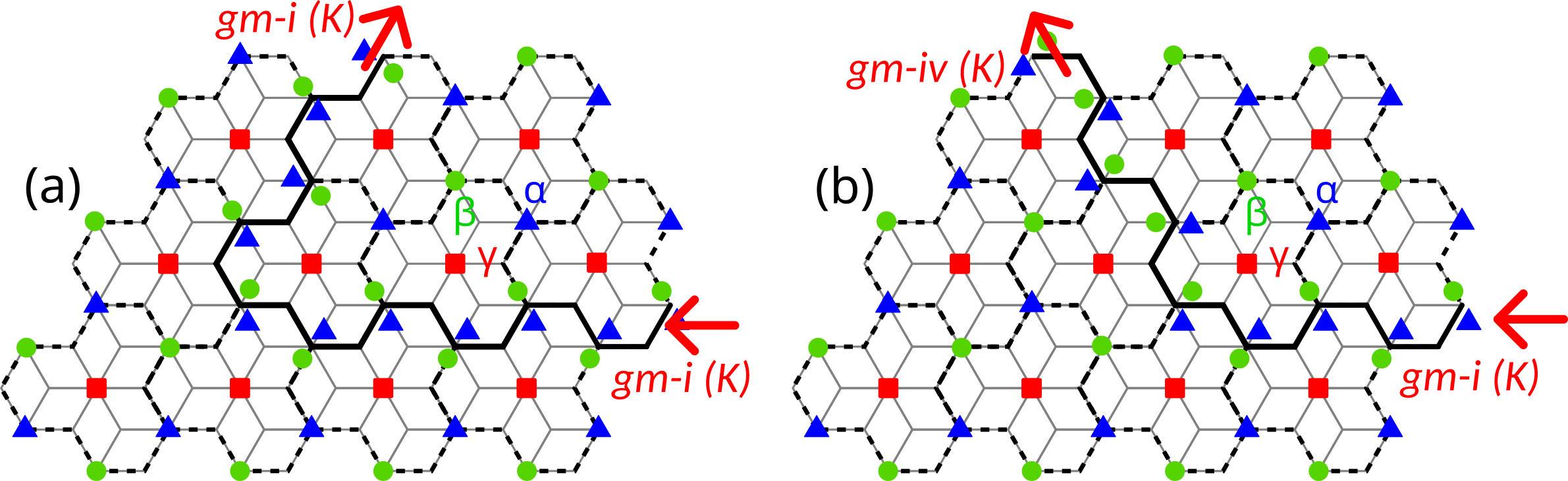}
	\caption{Relations between interfaces with 120°, (a), or 60°, (b), depending on the shape of the unit cell (dashed black line), for a \textit{gm-i} interface.}
	\label{fig:fig38}
\end{figure}
As shown in Fig.~\ref{fig:fig38} for a \textit{gm-i} interface, with the proper choice of unit cell, highlighted with a dashed line, a 120° corner conserves the geometry of the interface, while a 60° corner connects the \textit{gm-i} to the \textit{gm-iv} edge, where \textit{i} and \textit{iv} correspond to the interfaces in Fig.~\ref{fig:fig2}. Similarly, the \textit{gm-ii} interface can be connected to the \textit{gm-v} interface, and the \textit{gm-iii} to the \textit{gm-vi} interface. This is fully equivalent in a standard honeycomb lattice to connecting a large holes bearded interface with a small holes bearded interface in a 60° corner. Similarly, \textit{bm-i} and \textit{di-i} interfaces are connected respectively to \textit{bm-i} and \textit{di-i} interfaces for a 120° corner, but to \textit{bm-iv} and \textit{di-iv} interfaces in a 60° corner. Choosing other unit cells mixes glide-mirror and broken-mirror interfaces, see Fig.~S3, but we will not consider those cases in this work.

\begin{figure*}[t!]
	\includegraphics[width=16cm]{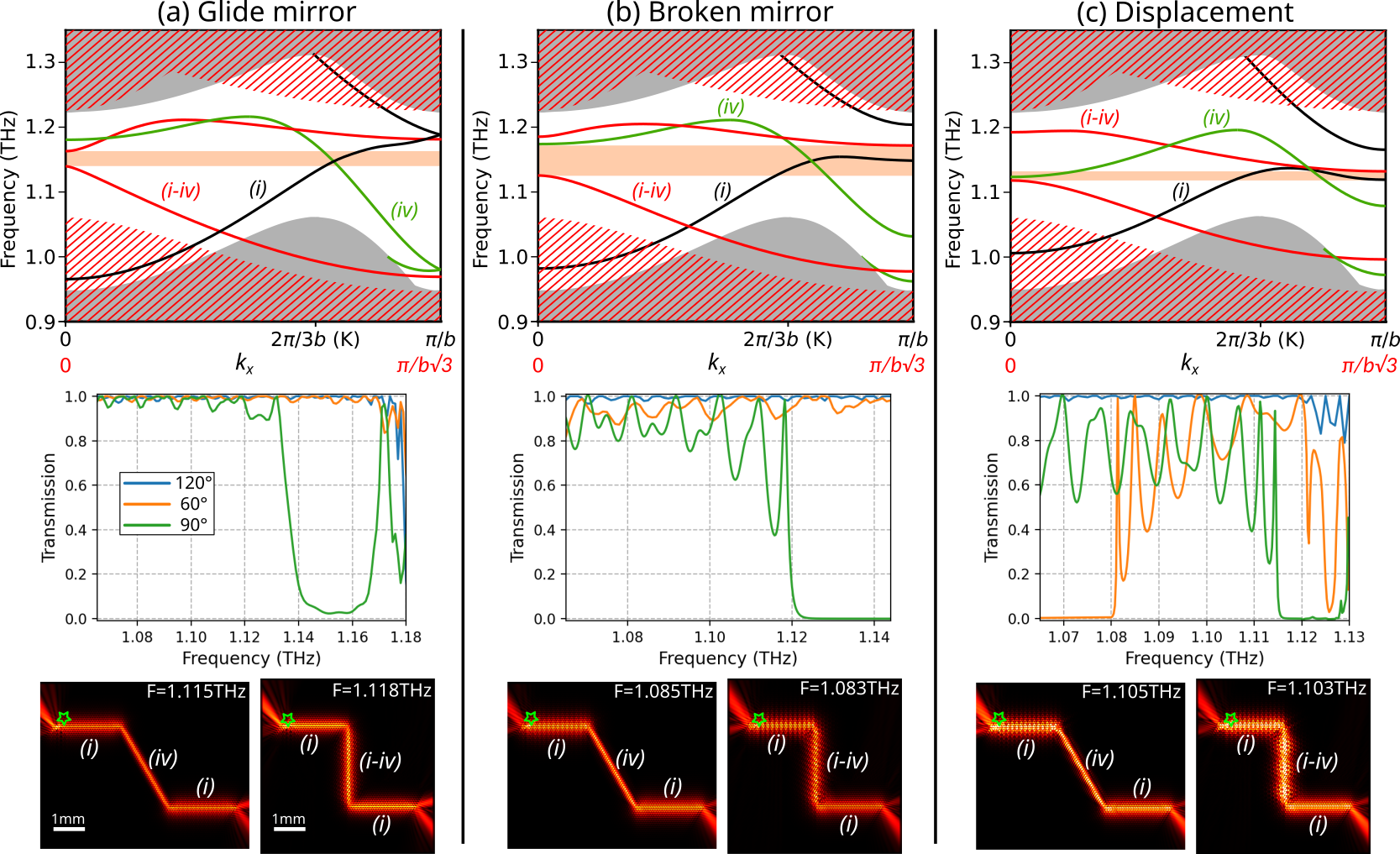}
	\caption{Transmission of edge-modes through 120°, 60° and 90° double corners for interfaces built on (i) and (iv) types for glide-mirror, (a), broken-mirror, (b), and displacement, (c), families. Top: dispersion curves of $\Gamma$K (black and green lines, $k_x\in\qty[0,\pi/b]$, projected bulk modes in gray) and $\Gamma$M (red lines, $k_x\in\qty[0,\pi/b/\sqrt{3}]$, projected bulk modes in red hatching) interfaces. Middle: transmission spectra. Bottom: distributions of magnetic-field amplitude for selected wavelengths.}
	\label{fig:fig33}
\end{figure*}
Figure \ref{fig:fig33} compares the propagation of edge-modes along 60° and 90° double-corners, only for combinations of interfaces (\textit{i}) and (\textit{iv}) in glide-mirror (a), broken-mirror (b), and displacement (c) families.
The dispersion relations are plotted in each case for $\Gamma$K and $\Gamma$M interfaces. The typical shape of a 90° corner and the super-cell with period $b\sqrt{3}$ used to compute the dispersion relations of the $\Gamma$M edge-modes are illustrated in Fig.~S4 of the Supplemental Materials. The label (\textit{i-iv}) for $\Gamma$M interfaces means that they can be seen as alternating interfaces of type (\textit{i}) and (\textit{iv}). Note that the $\Gamma$M edge-modes are gapped close to the frequency for which $\Gamma$K interfaces (\textit{i}) and (\textit{iv}) cross.

The computed transmission spectra for 120°, 60° and 90° double corners are displayed in the middle row of Fig.~\ref{fig:fig33}. Distributions of magnetic field amplitude for selected wavelengths for 60° and 90° corners are shown in the lower row of Fig.~\ref{fig:fig33} (see Figs.~S5 and S6 in Supplemental Materials for simulations of all other interfaces). Compared to 120° double corners, where all three edge modes travel without noticeable backscattering, in the case of 60° and 90° corners the considered interfaces show qualitative differences: indeed, only edge-modes propagating along glide-mirror symmetric interfaces, Fig.~\ref{fig:fig33}(a), show high transmission with small amount of back-reflection, both along 60° and 90° corners \textcolor{black}{(in which case the transmission drops by a maximum of 10\% beyond 1.12 THz)}. For this type of interface, in the case of a 60° corner, the high transmission can still be associated to topological arguments as, despite the geometrical intuition, \textcolor{black}{the propagation occurs along the same valley \cite{he_Investigation_2023}}. Indeed, such an angle connects the \textit{gm-i}, with positive group velocity, and the \textit{gm-iv}, with negative group velocity, interfaces. If we suppose that the \textit{gm-i} modes travel along the $\Gamma$K direction, the energy flows along the $\Gamma$K' directions for the \textit{gm-iv} edge modes but its wavevector is actually along the $\Gamma$K direction, \textcolor{black}{reversed as compared to the energy flow}. The large transmission along 90° corners can be interpreted based on the fact that this edge is actually a succession of short \textit{(i)} and \textit{(iv)} interfaces with alternating $\pm 60°$ angles, preserving then the valley number.

Transmission for broken-mirror interface drops on average by 10\% and 20\% respectively for 60° and 90° corners, while for displacement interfaces it drops by more than 40\% in both cases (see middle panels of Fig.~\ref{fig:fig33}(b), (c)). The presence of backreflection clearly appears in the form of interference patterns in the field distributions in the first and second segments of the waveguides (lower panels of Fig.~\ref{fig:fig33}(b), (c)).

\begin{figure*}[!t]
	\includegraphics[width=16cm]{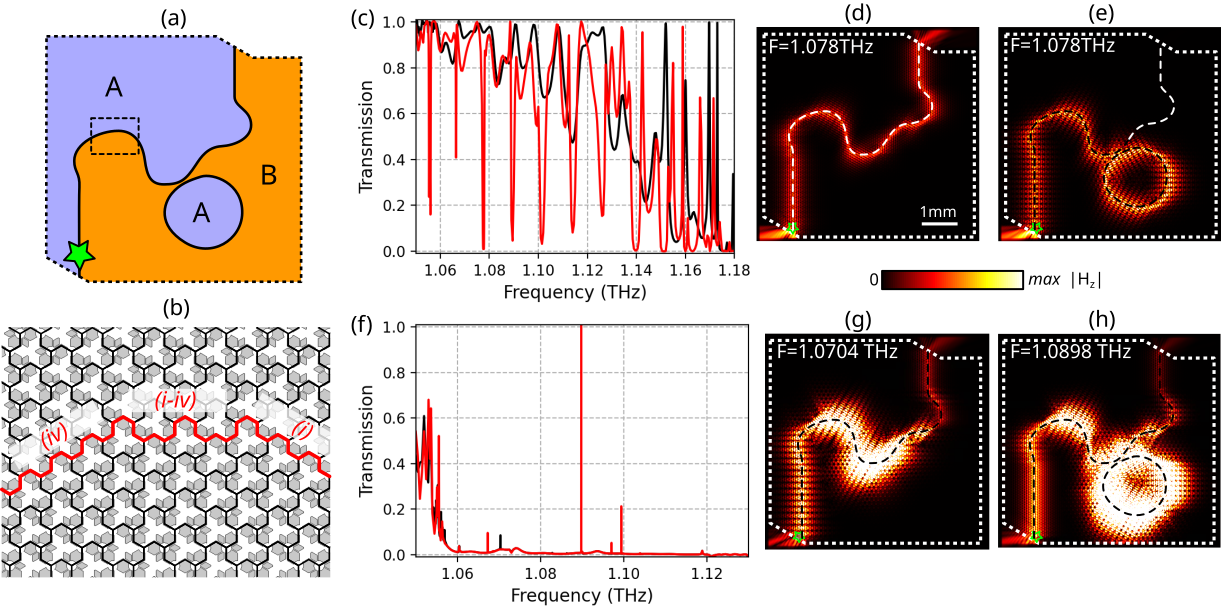}
	\caption{Transmission of edge-modes through a domain wall with an arbitrary shape between lattice $A$ and $B$, (a). The source is located at the position of the green star. (b): Detail of the interface in the area corresponding to the dashed box in (a). (c): Transmission spectra along the edge built on \textit{gm-i} and \textit{gm-iv} interfaces, without (black) and with (red) resonator. (d,e): Magnetic field amplitude distribution along the edge, respectively without and with resonator. (f-h): Same as (c-e) for displacement interfaces.}
	\label{fig:fig32}
\end{figure*}

Finally, we compare the transmission of edge-modes along a domain wall with arbitrary shapes for glide-mirror and displacement interfaces. We consider the interface plotted in Fig.~\ref{fig:fig32}(a), where $A$ is the original lattice and $B$ is its image. The shape of the interface has been constructed in such a way that on the length scale of a period it consists of a succession of short portions of alternating types of interfaces (\textit{i}) and (\textit{iv}), as shown in Fig.~\ref{fig:fig32}(b). This is ensured by the shape of the unit cells, highlighted with a thick black line centered on a sub-lattice rotation axis and preventing the mixing between glide-mirror and broken-mirror interfaces, see Fig.~S3. For both glide-mirror and displacement interfaces, we consider two cases: an interface waveguide coupled with an approximately round resonator (illustrated in Fig.~\ref{fig:fig32}(a)), and the same interface waveguide without the resonator. The results of the simulations for the glide-mirror edge are shown in Figs.~\ref{fig:fig32}(c-e). Transmissions along the domain wall without or with the cavity are plotted on Fig.~(c), respectively in black and red, with corresponding magnetic-field distributions shown in Figs.~(d) and (e). Figures (f) to (h) correspond to the displacement interfaces. The intermediate case of the broken-mirror domain-wall is shown in Fig.~S7 of Supplemental Materials.

The qualitative differences are striking between the two families of interfaces shown in Fig.~\ref{fig:fig32}. For the waveguide without resonator constructed from glide-mirror interfaces, the transmission stays relatively high (black line in Fig.~\ref{fig:fig32}(c)): above 50\% for a frequency between 1.06 THz and 1.13 THz, and above 75\% in the range 1.06-1.10 THz. It drops rapidly after 1.14 THz, close to the low limit of the $\Gamma$M band-gap in Fig.~\ref{fig:fig33}(a). The coupling to the cavity introduces a series of clear resonances visible in the red line of Fig.~\ref{fig:fig32}(c) with a frequency periodicity of about 12 GHz. A transmission minimum is observed at 1.078 THz, as plotted in Fig.~\ref{fig:fig32}~(e). These results are very similar to the case of simple straight waveguides coupled to triangular resonators investigated above, with the difference that for the configuration of Fig.~\ref{fig:fig32}, weak backscattering occurs all along the waveguide. This is clearly visible as interference patterns in movies M1 and M2 in Supplemental Materials. Such distributed backscattering events were expected as the edge now does not show any local translation symmetry on a distance larger than a few periods, and simulations for corners different from 120° show weak but non-zero backscattering even for glide-mirror interfaces.

The simulations for the edge constructed from displacement interfaces are displayed in Fig.~\ref{fig:fig32}(f)-(h). Contrary to the glide-mirror interface, the transmission is very low except for a single frequency close to 1.090 THz where a sharp resonance with transmission close to unity is observed. This isolated event is related to random coherent back-scattering along the edge leading to destructive interferences of the reflected edge-mode. The average low transmission originates in the larger back-reflection already observed along 60° and 90° double-corners for displacement interfaces, as plotted in Fig.~\ref{fig:fig33}(c). Interestingly, the waveguide built on broken-mirror interfaces (see Supplemental Materials) shows intermediate behavior, with a low level of transmission but a larger number of narrow resonances with higher transmission than for the displacement edge (Fig.~S7).

\section{Conclusions}
Those results evidence that the robustness of the propagation of electromagnetic waves along domain walls separating valley topological photonic crystals is not uniquely determined by the topological figures of the bulk (Berry curvature and valley Chern numbers), but also by the symmetry of the interface and the shape of the domain wall on a length-scale on the order of a few periods.
A striking conclusion of our work is that robust propagation along 120° shape corners is not necessarily related to a valley topological phase transition at the interface, as an edge constructed between two valley photonic crystals differing only by a translation (with identical valley Chern number on both sides of the interface but non zero Berry curvature) leads to as robust propagation as canonical interfaces built on glide-mirror image VPCs. In the latter case, robust propagation is attributed to valley number conservation, which is a global property of the bulk lattice, while in the former case, the helicity of the edge-mode propagating along a chiral interface is conserved, which is a property of the interface (in particular, glide-mirror interfaces are not chiral). However, along domain walls composed of long (many periods) linear parts connected with 60° or 90° corners, it appears that an actual topological phase transition is a necessary condition to ensure robustness, related to the conservation of the valley number despite connecting locally different interfaces. Even if this condition is ensured, the robustness seems to be lower if the glide-mirror symmetry is broken, in our case by an additional sub-lattice translation of the image VPC, corresponding to the broken-mirror interfaces in our work. Finally, in the extreme case where the domain wall takes an arbitrary shape, which means that it is built from a succession of short (down to a few periods) interfaces connected with 120°, 60° or 90° angles, the local symmetry of the edge has to be a glide-mirror in order to reach appreciable transmission coefficients, however lower than unity.

\textcolor{black}{Our work, together with recent publications \cite{yang_Evolution_2021,torres_canonical_2024a}, tends to question the relation between the valley topology of the bulk lattice and the robustness of the propagation, particularly in circuits built on a strict succession of 120° corners, but however establishes a clear hierarchy between more complex domain walls built at interfaces separating lattices with inversion of the Berry curvature and those without, while in the same time confirming the strong role of the geometry and symmetry of the interfaces on the long and short length-scale.} Due to the richness of the investigated lattice, this work has been restricted to a few but meaningful types of interfaces which locally correspond to bearded interfaces in honeycomb valley photonic crystals with two sites per unit-cell. Despite not being fully exhaustive, we believe that the conclusions of our work give an original lighting on the relation between interface geometry and robust propagation of electromagnetic edge modes, necessary to the development of efficient and compact photonic platforms.

\section*{Acknowledgements}
G.L. and Y.P. are grateful for the Horizon-RIA action project Magnific (Project No. 101091968). A.A. is grateful to the European Research Council for support via the
grant EmergenTopo (Grant No. 865151).

\section*{Data availability}
The data that support the findings of this study are available from the corresponding author upon reasonable request.


%

\newpage

\foreach \x in {1,...,8}
{%
	\clearpage
	\includepdf[pages={\x}]{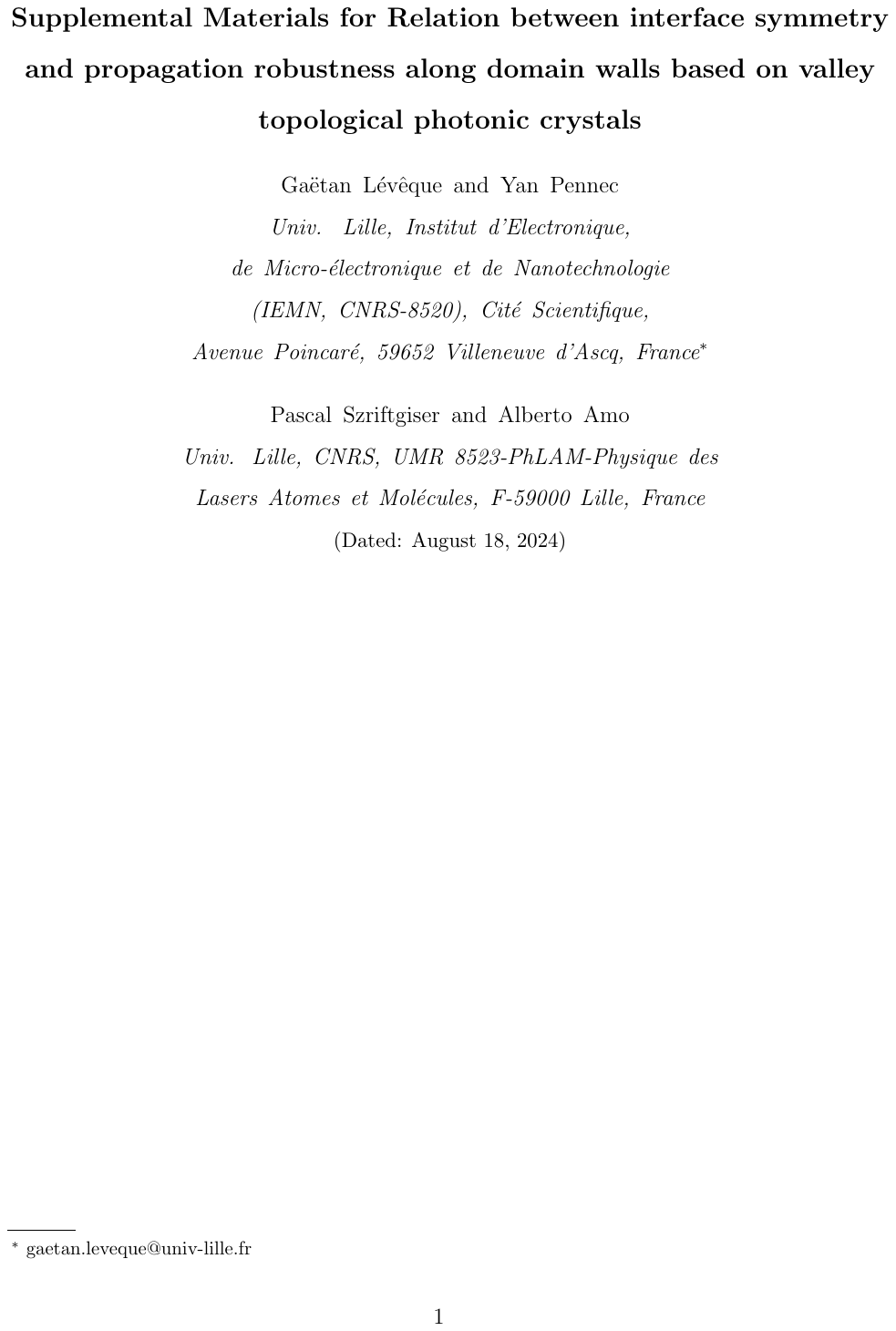} 
}

\end{document}